\documentclass[]{aastex631}

\begin{document}

\title{Comment on the feasibility of carbon burning in Betelgeuse: a response to ``The evolutionary stage of Betelgeuse inferred from its pulsation periods", arXiv:2306.00287}

\author[0000-0002-8159-1599]{L\'{a}szl\'{o} Moln\'{a}r}
\email{molnar.laszlo@csfk.org}
\affiliation{Konkoly Observatory, ELKH CSFK, Budapest, Konkoly Thege Mikl\'os \'ut 15-17, Hungary}
\affiliation{E\"otv\"os Lor\'and University, Institute of Physics, H-1117 P\'azm\'any P\'eter s\'et\'any 1/A, Budapest, Hungary}

\author[0000-0002-8717-127X]{Meridith Joyce}
\email{meridith.joyce@csfk.org}
\affiliation{Konkoly Observatory, ELKH CSFK, Budapest, Konkoly Thege Mikl\'os \'ut 15-17, Hungary}

\author[0000-0002-4972-3803]{Shing-Chi Leung}
\email{leungs@sunypoly.edu}
\affiliation{Department of Mathematics and Physics, SUNY Polytechnic Institute, 100 Seymour Road, Utica NY 13502, USA}

\begin{abstract}
The recent pre-print\footnote{version v1 published to arxiv on June 3rd, 2023.} by \citet{Saio2023} argues that the supergiant Betelgeuse is already undergoing carbon burning, based on the assumption that all of its light variations are caused by radial pulsations. However, the angular diameter measurements of the star are in conflict with the stellar radius required by their models, as we show in this note. We 
discuss
the feasibility that the Great Dimming was caused by constructive mode interference using long-term brightness measurements and comment on differences in modeling frameworks adopted in \citet{Saio2023} vs \citet{Joyce2020}. 

\end{abstract}


\section{Introduction} \label{sec:intro}
A recent pre-print by \citet{Saio2023} postulates that Betelgeuse is already in the carbon-burning phase and is therefore close to a supernova explosion. 
Crucially, they set the constraint that the Long Secondary Period (LSP) in Betelgeuse's light variations is in fact the fundamental radial mode (FM). LSPs are observed in many red giants and supergiants and are generally connected to external processes like dust and/or binarity \citep[see, e.g.,][]{wood-nicholls-2009,soszynski-2021}, although internal processes like pulsation \citep{Derekas-2006,Saio-2015} have also been proposed.

The light variation of Betelgeuse is complex and comprises multiple signals.
Usually the longest periodicity, at 2200 days, is considered to be an LSP with an unknown origin, whereas the 380-430 d period is identified as the fundamental radial \textit{p}-mode. Further signals, such as the 185 d periodicity identified by \citet{Joyce2020} and discussed in \citet{dupree-2022} and \citet{MacLeod2023}, are identified as overtones. \citet{Saio2023} instead propose that all signals are \textit{p}-mode pulsations, and therefore the 2200 d period is the fundamental mode. This requires a much larger stellar radius---on the order of 1400 $R_{\odot}$---than those inferred by other studies, between 600--1000 $R_\odot$ \citep{Dolan-2016}. 

\section{Comments on observational constraints}
The diameter of Betelgeuse has been measured in several wavelengths. However, measurements do not necessarily detect the visible-light photosphere of the star directly. They are affected by factors like limb darkening, spots, molecular layers and circumstellar dust. 
Therefore the raw numbers cannot be used as a constraint for the stellar radius. \citet{Haubois-2009} found a diameter of 44.3--46.74\,mas in the \textit{H}-band depending on modeling assumptions. \citet{montarges-2014} found a limb-darkened diameter of 42.28$\pm$0.43 mas, using continuum \textit{K}-band observations that are the least affected by molecular layers. Other studies using near-infrared observations also place the photospheric diameter to between 42--45\,mas \citep[][and references therein]{Dolan-2016}. 

The picture gets more complicated in the mid-infrared. As \citet{Cannon-2023} explains, the apparent stellar diameter peaks at around 11\,$\mu m$, where the contribution of hot circumstellar dust is the greatest. They specifically state that given the complexities in the mid-infrared emission of the star, ``values for the disk diameter from $\lambda > 8.75\, \mu m$, should not be taken at face value.'' 

We conclude therefore that the angular diameter of the visible-light photosphere of Betelgeuse has been constrained, by multiple authors, to be below 45 mas. This limits the physical radius to $R < 1100\, R_\odot$ even when using the farthest distance value of 222\,pc \citep{harper-2017}, and gives greater validity to sub--1000\,$R_\odot$ values based on 
most other distance measurements and inferences, some of which are summarized in Figure 10 of \citet{Joyce2020}. These are in direct conflict with the radius required for a 2200-d pulsation period.
 
\begin{figure*}
\centering
\includegraphics[width=\textwidth]{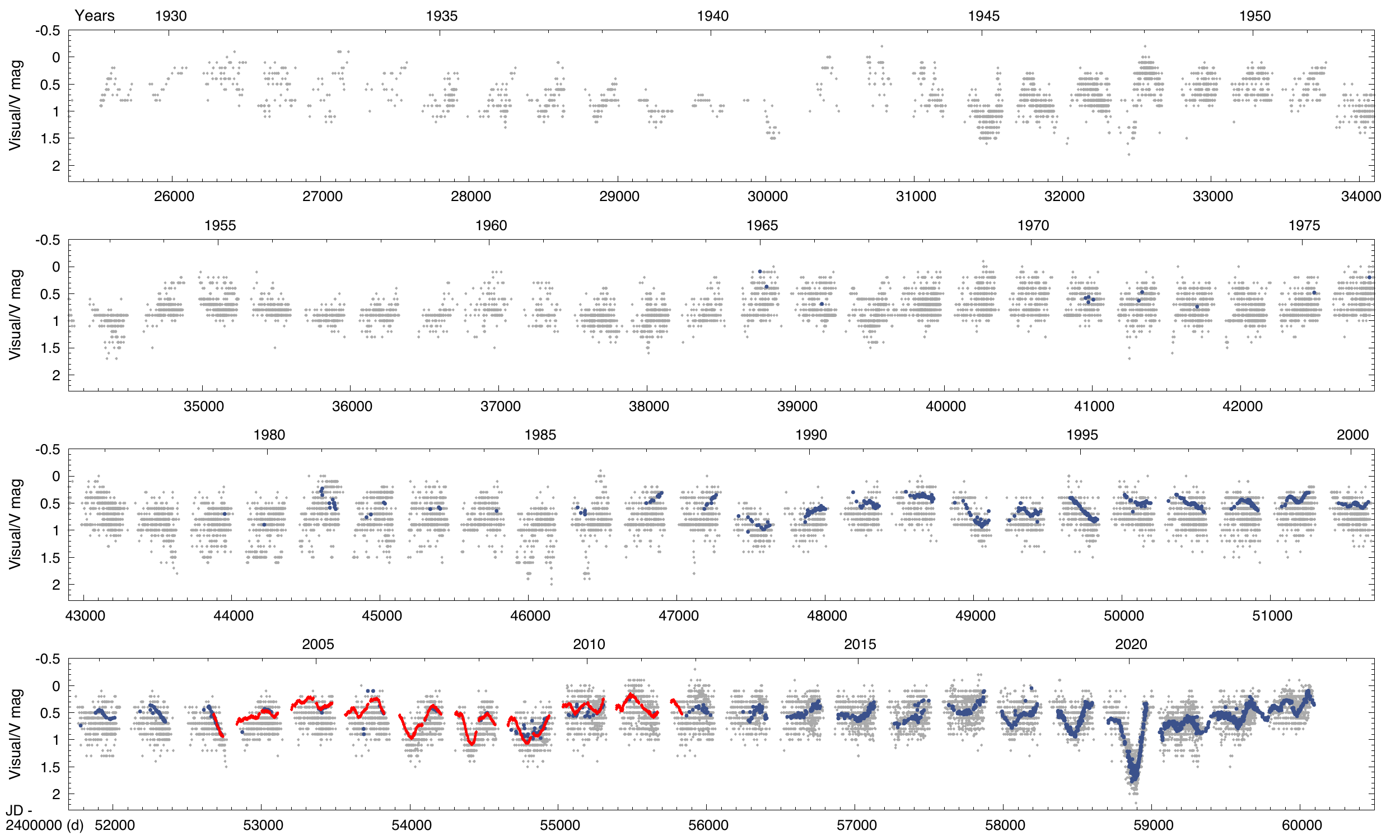}
\caption{Visual (grey) and photometric (blue: \textit{V}-band; red: SMEI) light curves of Betelgeuse. Updated from Fig.~1 in \citet{Joyce2020}. }
\label{fig:lightcurve}
\end{figure*}

In Fig.~\ref{fig:lightcurve} we update the light curve published in \citet{Joyce2020} with new measurements. We can see the slow undulation caused by the LSP, but it is clear that the dimming far exceeded the extrema of its usual brightness variations. 
The LSP has always been present, and it was  
pronounced in the 1930s to 1960s. The long-term light curve therefore does not support the hypothesis that the 2200-d variation has been increasing recently and that the cause of the Great Dimming was constructive mode interference. That would cause a more even variation between cycles, as the synthesized light curves of \citet{Saio2023} themselves suggest. Instead, a sudden dimming can be naturally explained by the condensation of a dense dust cloud in our line of sight \citep{dust-cloud-2021Natur.594..365M}.

\section{Comments on modeling}

\citet{Saio2023} attempt to explain the cause of the Great Dimming 
using a fully linear theoretical framework. 
Their models, like all of those computed in \citet{Joyce2020}, do not capture behavior beyond the outer physical boundary of the star, yet many have argued that dust-gas interactions are necessary considerations. The need to invoke constructive interference of modes only exists if the cause of the Great Dimming was intrinsic, 
yet there is little ambiguity that the Great Dimming was caused by a dust cloud \citep{dust-cloud-2021Natur.594..365M}.

The right-hand panel of Figure 12 in \citet{Joyce2020} shows a synthetic period-radius relationship based on extracted cycle lengths from 1D hydrodynamic simulations (MESA). This relation does indeed associate an FM period of 2200 d to a much larger radius: roughly $1600 R_{\odot}$. 

Likewise, the linear, adiabatic seismic calculations computed with GYRE in \citet{Joyce2020} are shown in terms of O1/FM ratio in their Figure 9. Based on this plot, \citet{Saio2023}'s O1/FM ratio of 0.19 
would imply a present-day mass of more than $24 M_{\odot}$. While there are structural differences between helium- and carbon-burning models, these differences are interior, whereas the pressure modes probe the convective envelope.

Key differences in the modeling formalisms are as follows:  

\begin{itemize}
    \item \citet{Saio2023} calculate the perturbation of luminosity using (Eq. 2) $\delta L/L = 4 \delta T_{\rm eff}/T_{\rm eff} + 2 \delta R/R$, which only applies in the condition that the photosphere is fixed, which is not the case for the extended and dilute atmosphere. 
    
    \item The non-adiabatic calculations in \cite{Saio2023} do not capture the non-linear density and temperature dependence of opacity, thus neglecting the complication of kappa mechanism and its mode excitation.
    
    \item They use the non-adiabatic oscillation which linearly extends the adiabatic radial oscillation and includes heat transport perturbation from convection, whereas the \citet{Joyce2020} models are in fully consistent hydrodynamics.

    \item \citet{Joyce2020} used hydrodynamics to model initial amplitude growth. \citet{Saio2023} 
    assumed a mode amplitude growth profile.

\end{itemize}

Nonetheless, it is our stance that
the main 
source of tension between our earlier work and \citet{Saio2023}
stems from their treating the LSP as radial pulsation and applying inadequate observational constraints to justify that assumption.

\begin{acknowledgments}
\textit{Acknowledgements:} This research was supported by the `SeismoLab' KKP-137523 \'Elvonal grant of the NKFIH. M.J.\ acknowledges funding 
from the European Union's Horizon 2020 research and innovation programme. Twitter discussions with Dr.\ Miguel Montar\'es, Dr.\ Thayne Currie and Prof.\ Chris Lintott are gratefully acknowledged. 
\end{acknowledgments}

\vspace{5mm}
\facilities{AAVSO}


\bibliography{betelgeuse_note}{}

\begin{thebibliography}{}
\expandafter\ifx\csname natexlab\endcsname\relax\def\natexlab#1{#1}\fi
\providecommand{\url}[1]{\href{#1}{#1}}
\providecommand{\dodoi}[1]{doi:~\href{http://doi.org/#1}{\nolinkurl{#1}}}
\providecommand{\doeprint}[1]{\href{http://ascl.net/#1}{\nolinkurl{http://ascl.net/#1}}}
\providecommand{\doarXiv}[1]{\href{https://arxiv.org/abs/#1}{\nolinkurl{https://arxiv.org/abs/#1}}}

\bibitem[{{Cannon} {et~al.}(2023){Cannon}, {Montarg{\`e}s}, {de Koter},
  {Matter}, {Sanchez-Bermudez}, {Norris}, {Paladini}, {Decin}, {Sana},
  {Sundqvist}, {Lagadec}, {Kervella}, {Chiavassa}, {Dupree}, {Perrin},
  {Scicluna}, {Stee}, {Kraus}, {Danchi}, {Lopez}, {Millour}, {Drevon},
  {Cruzal{\`e}bes}, {Berio}, {Robbe-Dubois}, \& {Rosales-Guzman}}]{Cannon-2023}
{Cannon}, E., {Montarg{\`e}s}, M., {de Koter}, A., {et~al.} 2023, arXiv
  e-prints, arXiv:2303.08892, \dodoi{10.48550/arXiv.2303.08892}

\bibitem[{{Derekas} {et~al.}(2006){Derekas}, {Kiss}, {Bedding}, {Kjeldsen},
  {Lah}, \& {Szab{\'o}}}]{Derekas-2006}
{Derekas}, A., {Kiss}, L.~L., {Bedding}, T.~R., {et~al.} 2006, \apjl, 650, L55,
  \dodoi{10.1086/508686}

\bibitem[{{Dolan} {et~al.}(2016){Dolan}, {Mathews}, {Lam}, {Quynh Lan},
  {Herczeg}, \& {Dearborn}}]{Dolan-2016}
{Dolan}, M.~M., {Mathews}, G.~J., {Lam}, D.~D., {et~al.} 2016, \apj, 819, 7,
  \dodoi{10.3847/0004-637X/819/1/7}

\bibitem[{{Dupree} {et~al.}(2022){Dupree}, {Strassmeier}, {Calderwood},
  {Granzer}, {Weber}, {Kravchenko}, {Matthews}, {Montarg{\`e}s}, {Tappin}, \&
  {Thompson}}]{dupree-2022}
{Dupree}, A.~K., {Strassmeier}, K.~G., {Calderwood}, T., {et~al.} 2022, \apj,
  936, 18, \dodoi{10.3847/1538-4357/ac7853}

\bibitem[{{Harper} {et~al.}(2017){Harper}, {Brown}, {Guinan}, {O'Gorman},
  {Richards}, {Kervella}, \& {Decin}}]{harper-2017}
{Harper}, G.~M., {Brown}, A., {Guinan}, E.~F., {et~al.} 2017, \aj, 154, 11,
  \dodoi{10.3847/1538-3881/aa6ff9}

\bibitem[{{Haubois} {et~al.}(2009){Haubois}, {Perrin}, {Lacour}, {Verhoelst},
  {Meimon}, {Mugnier}, {Thi{\'e}baut}, {Berger}, {Ridgway}, {Monnier},
  {Millan-Gabet}, \& {Traub}}]{Haubois-2009}
{Haubois}, X., {Perrin}, G., {Lacour}, S., {et~al.} 2009, \aap, 508, 923,
  \dodoi{10.1051/0004-6361/200912927}

\bibitem[{{Joyce} {et~al.}(2020){Joyce}, {Leung}, {Moln{\'a}r}, {Ireland},
  {Kobayashi}, \& {Nomoto}}]{Joyce2020}
{Joyce}, M., {Leung}, S.-C., {Moln{\'a}r}, L., {et~al.} 2020, \apj, 902, 63,
  \dodoi{10.3847/1538-4357/abb8db}

\bibitem[{{MacLeod} {et~al.}(2023){MacLeod}, {Antoni}, {Huang}, {Dupree}, \&
  {Loeb}}]{MacLeod2023}
{MacLeod}, M., {Antoni}, A., {Huang}, C.~D., {Dupree}, A., \& {Loeb}, A. 2023,
  arXiv e-prints, arXiv:2305.09732, \dodoi{10.48550/arXiv.2305.09732}

\bibitem[{{Montarg{\`e}s} {et~al.}(2014){Montarg{\`e}s}, {Kervella}, {Perrin},
  {Ohnaka}, {Chiavassa}, {Ridgway}, \& {Lacour}}]{montarges-2014}
{Montarg{\`e}s}, M., {Kervella}, P., {Perrin}, G., {et~al.} 2014, \aap, 572,
  A17, \dodoi{10.1051/0004-6361/201423538}

\bibitem[{{Montarg{\`e}s} {et~al.}(2021){Montarg{\`e}s}, {Cannon}, {Lagadec},
  {de Koter}, {Kervella}, {Sanchez-Bermudez}, {Paladini}, {Cantalloube},
  {Decin}, {Scicluna}, {Kravchenko}, {Dupree}, {Ridgway}, {Wittkowski},
  {Anugu}, {Norris}, {Rau}, {Perrin}, {Chiavassa}, {Kraus}, {Monnier},
  {Millour}, {Le Bouquin}, {Haubois}, {Lopez}, {Stee}, \&
  {Danchi}}]{dust-cloud-2021Natur.594..365M}
{Montarg{\`e}s}, M., {Cannon}, E., {Lagadec}, E., {et~al.} 2021, \nat, 594,
  365, \dodoi{10.1038/s41586-021-03546-8}

\bibitem[{{Saio} {et~al.}(2023){Saio}, {Nandal}, {Meynet}, \&
  {Ekst{\"o}m}}]{Saio2023}
{Saio}, H., {Nandal}, D., {Meynet}, G., \& {Ekst{\"o}m}, S. 2023, arXiv
  e-prints, arXiv:2306.00287.
\newblock \doarXiv{2306.00287}

\bibitem[{{Saio} {et~al.}(2015){Saio}, {Wood}, {Takayama}, \&
  {Ita}}]{Saio-2015}
{Saio}, H., {Wood}, P.~R., {Takayama}, M., \& {Ita}, Y. 2015, \mnras, 452,
  3863, \dodoi{10.1093/mnras/stv1587}

\bibitem[{{Soszy{\'n}ski} {et~al.}(2021){Soszy{\'n}ski}, {Olechowska},
  {Ratajczak}, {Iwanek}, {Skowron}, {Mr{\'o}z}, {Pietrukowicz}, {Udalski},
  {Szyma{\'n}ski}, {Skowron}, {Gromadzki}, {Poleski}, {Koz{\l}owski}, {Wrona},
  {Ulaczyk}, \& {Rybicki}}]{soszynski-2021}
{Soszy{\'n}ski}, I., {Olechowska}, A., {Ratajczak}, M., {et~al.} 2021, \apjl,
  911, L22, \dodoi{10.3847/2041-8213/abf3c9}

\bibitem[{{Wood} \& {Nicholls}(2009)}]{wood-nicholls-2009}
{Wood}, P.~R., \& {Nicholls}, C.~P. 2009, \apj, 707, 573,
  \dodoi{10.1088/0004-637X/707/1/573}

\end{thebibliography}
\bibliographystyle{aasjournal}

\end{document}